\documentclass[11pt,twoside]{article}
\usepackage{asp2014}

\aspSuppressVolSlug
\resetcounters

\begin{document}

\title{Testing Fundamental Particle Physics with the
Galactic White Dwarf Luminosity Function}
\author{M. M. Miller Bertolami,$^{1,2,\dagger}$ B. E. Melendez,$^{2,3}$, L. G. Althaus$^{2,3}$ and J. Isern$^{4,5}$
\affil{$^1$Max-Planck-Institut f\"ur Astrophysik, Karl-Schwarzschild-Str. 1, 85748, Garching, Germany; \email{marcelo@MPA-Garching.MPG.DE}}
\affil{$^2$Instituto de Astrof\'isica de La Plata, UNLP-CONICET, Paseo del
  Bosque s/n, 1900 La Plata, Argentina; \email{mmiller@fcaglp.unlp.edu.ar}}
\affil{$^3$Grupo de evoluci\'on estelar y pulsaciones, Facultad de Ciencias
  Astron\'omicas y Geof\'isicas, Universidad Nacional de La Plata, Paseo del
  Bosque s/n, 1900 La Plata, Argentina; \email{brenmele@gmail.com, althaus@fcaglp.unlp.edu.ar}}
\affil{$^4$Institut de Ci\'encies de l'Espai (CSIC), Facultat de  Ci\'encies, Campus UAB, Torre C5-parell, 08193 Bellaterra, Spain}
\affil{$^5$Institute for Space Studies of Catalonia, c/Gran Capit\'a
 2-4,  Edif. Nexus 104, 08034 Barcelona, Spain; \email{isern@ieec.cat}}
\affil{$^\dagger$Postdoctoral fellow of the Alexander von Humboldt Foundation}
}
\paperauthor{M. M. Miller Bertolami}{marcelo@MPA-Garching.MPG.DE}{}{Max-Planck-Institut f\"ur Astrophysik}{Author1 Department}{Garching}{Bayern}{85748}{Germany}
\paperauthor{B. E. Melendez}{brenmele@gmail.com}{}{Instituto de Astrof\'isica
  de La Plata}{UNLP-CONICET}{La Plata}{Buenos Aires}{1900}{Argentina}
\paperauthor{L. G. Althaus}{althaus@fcaglp.unlp.edu.ar}{}{Instituto de Astrof\'isica  de La Plata}{UNLP-CONICET}{La Plata}{Buenos Aires}{1900}{Argentina}
\paperauthor{J. Isern}{isern@ieec.cat}{}{Facultat de  Ci\'encies, Campus UAB}{Institut de Ci\'encies de l'Espai (CSIC)}{Barcelona}{Catalunya}{08034}{Spain}

\begin{abstract}
Recent determinations of the white dwarf luminosity function (WDLF) from very
large surveys have extended our knowledge of the WDLF to very high
luminosities. It has been shown that the shape of the luminosity function of
white dwarfs (WDLF) is a powerful tool to test the possible properties and
existence of fundamental weakly interacting subelectronvolt particles. This,
together with the availability of new full evolutionary white dwarf models
that are reliable at high luminosities, have opened the possibility of testing
particle emission in the core of very hot white dwarfs. We use the available
WDLFs from the Sloan Digital Sky Survey and the SuperCOSMOS Sky Survey to
constrain the values of the neutrino magnetic dipole moment ($\mu_\nu$) and
the axion-electron coupling constant ($g_{ae}$) of DFSZ-axions.
\end{abstract}

\section{Introduction}
Astrophysical arguments can provide powerful constraints to the properties of
elementary particles. This is particularly true at the subelectronvolt scale
\citep{2010ARNPS..60..405J}. Weakly interacting subelectronvolt particles can
be abundantly created in the hot and dense stellar plasmas and then escape
from the stellar interior without further interactions (see
\citealt{1996slfp.book.....R} for a detailed review). Consequently, if they
exist, these particles provide a local energy sink for the stellar
structure. This alters the structural and evolutionary properties predicted
for different stars. The observable impact of these changes provide some of
the most powerful limits on the properties of real or hypothetical
subelectronvolt particles like neutrinos and axions
\citep{2013PhRvL.111w1301V}.  In particular, white dwarf cooling is sensitive
to the hypothetical existence of DFSZ-axions\footnote{DFSZ-axion models (after
  \citealt{1981PhLB..104..199D, Zhitnitsky}) allow for the coupling of axions
  to leptons.} or a magnetic dipole moment of the neutrino ($\mu_\nu$). This
allows using white dwarfs to constrain the values of the axion-electron
coupling constant ($g_{ae}=2.8\times 10^{-14}\times m_a^{\rm meV}\cos^2\beta$,
where $m_a$ is the mass of the axion) and $\mu_\nu$
\citep{1994MNRAS.266..289B}. In the context of the strong CP problem,
\cite{2008ApJ...682L.109I} have shown that modern white dwarf luminosity
functions (WDLF), coming from large sky surveys, offer a new possibility to
learn about elementary particle physics. Here we study the constraints that
can be derived for $\mu_\nu$ and $g_{ae}$ from the use of recent
determinations of the WLDF and the aid of state-of-the-art white dwarf models.

\section{The white dwarf luminosity function}
To construct the theoretical WDLFs to be compared with the derivations
of the WDLF of the Galactic disk we use the method described by
\cite{1989ApJ...341..312I}. In this approach the number of white dwarfs
per logarithmic luminosity and volume is computed as
\begin{equation}
\frac{dn}{dl}=-\int_{M_1}^{M_2} \psi(t) \left(\frac{dN}{dM}\right)
 \left(\frac{\partial t_c}{\partial l}\right)_m dM
\label{eq:dndl}
\end{equation}
where $\psi(t)$ is the galactic stellar formation rate at time $t$, $N(M)$ is
the initial mass function and $t_c(l,m)$ is the time since the formation of a
white dwarf, of mass $m$, for the star to reach a luminosity
$\log(L/L_\odot)=l$. In order to compute the integral in equation \ref{eq:dndl}
we also need the initial-final mass relation $m(M)$, and the pre-white dwarf
stellar lifetime $t_{ev}(M)$. We adopt a Salpeter initial mass function, the
$m(M)$ relation form \cite{2009ApJ...692.1013S} and the $t_{ev}(M)$ from the
BaSTI database\footnote{\tt http://basti.oa-teramo.inaf.it/}. It is worth noting
that, for a given white dwarf luminosity ($l$) and mass of the progenitor
($M$) the formation time of the star, $t$, is obtained by solving
$t+t_{ev}(M)+t_c(l,m)=T_{OS}$,
where $T_{OS}$ is the assumed age of the oldest star in the computed
population. The lowest initial mass that produces a white dwarf with
luminosity $l$ at the present time ($M_1$) is obtained from
$t_{ev}(M)+t_c(l,m)=T_{OS}$. The value of $M_2$ corresponds to the largest
stellar mass progenitor that produces a white dwarf. In what follows the value
of $\psi(t)$ is assumed constant and its value is obtained from normalization to
the observational WDLF.

In order to compute theoretical WDLFs that take advantage of new available
data that extends to the high luminosity regime, full evolutionary models
derived from the progenitor history are to be preferred. The initial white
dwarf models adopted in our simulations were taken from
\cite{2010ApJ...717..183R}. Then we computed the values of $t_c(l,m)$ under
the assumption of different axion masses and values of the magnetic dipole
moment of the neutrino by using {\tt LPCODE} stellar evolution
code. This was performed by including self-consistently the anomalous energy
losses in the computation of the cooling sequences. See
\cite{2014A&A...562A.123M} and \cite{2014arXiv1406.7712M} for details.
Sequences were computed for axion masses of $m_a$ $\cos^2\beta$= 2.5, 5, 7.5,
10, 15, 20 \& 30 meV and neutrino magnetic dipole moments of $\mu_{12}$ = 1,
2, 5 and 10, where $\mu_{12}=\mu_\nu/(10^{-12} e\hbar/(2m_e
c))$, as well as for the standard case ($\mu_{12}=m_a=0$). 



\begin{figure}[ht]
\includegraphics[clip, angle=0, width=13.cm]{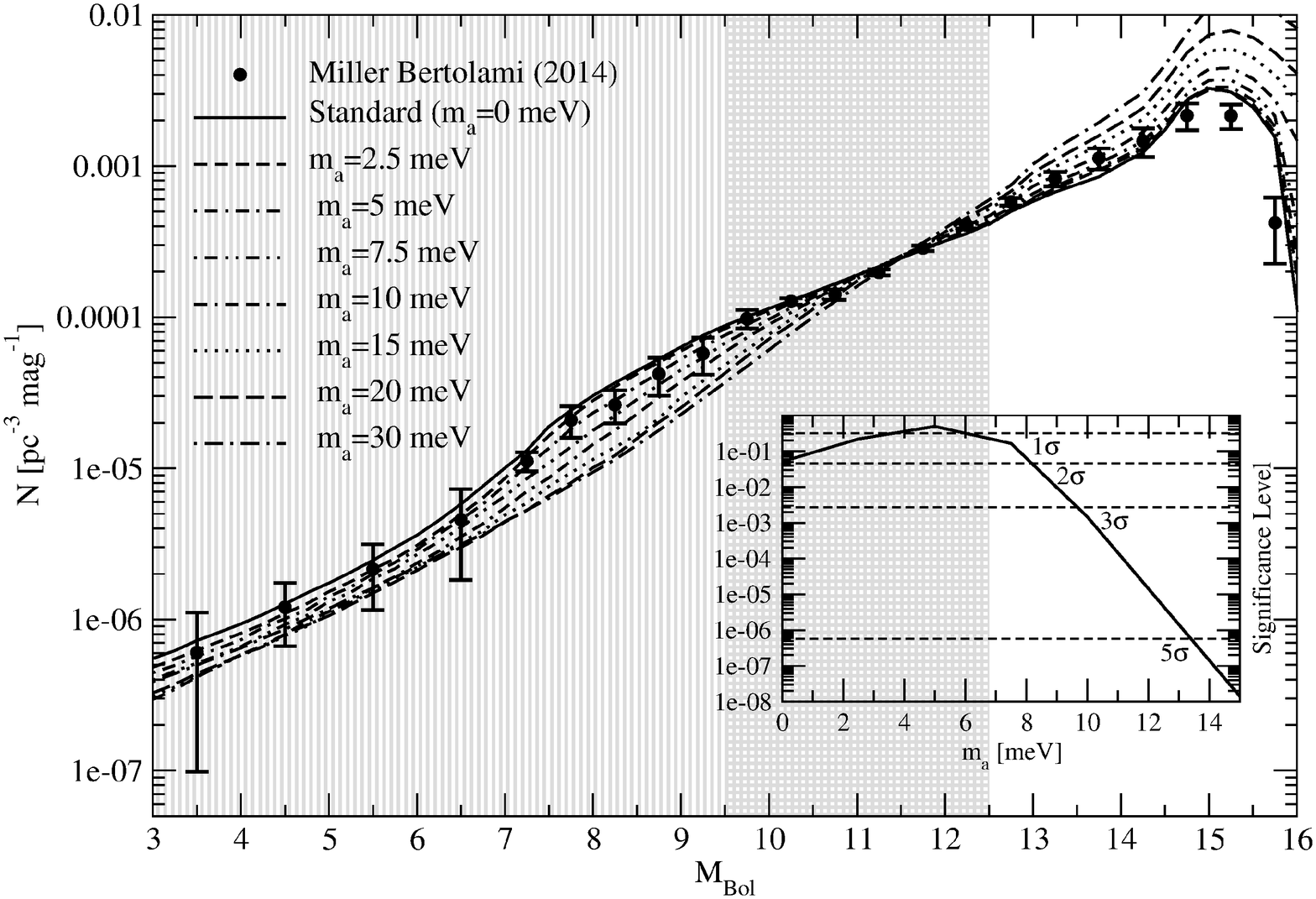} 
\includegraphics[clip, angle=0, width=13.cm]{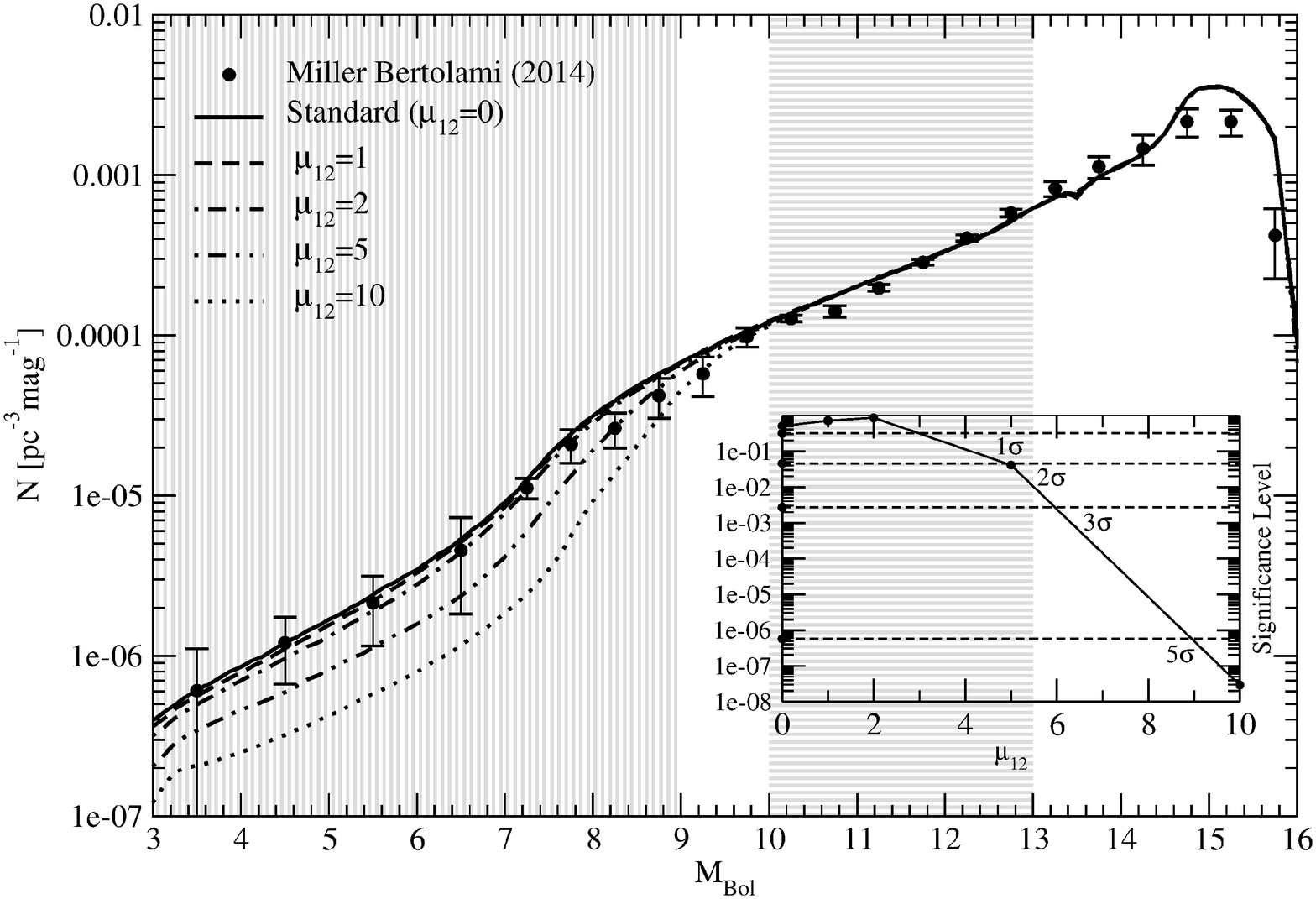} 
\caption{{\it Upper Panel:} Comparison of the WDLF estimated by \cite{2014A&A...562A.123M} (from the
  independent WDLFs of \citealt{2006AJ....131..571H},
  \citealt{2009A&A...508..339K} and \citealt{2011MNRAS.417...93R}) with the
  theoretical WDLFs computed under the assumption of different masses for the
  DFSZ-axion. Vertical grey lines indicate the range adopted for the
  comparison in the $\chi^2$-test, while horizontal grey lines indicate the
  range of luminosities adopted for the normalization of the theoretical
  WDLFs. {\it Lower Panel:} Same as the upper panel but for the theoretical
  WDLFs derived under different assumptions for the value of the magnetic
  dipole moment of the neutrino. Insets show the significance levels for
  $\chi^2$-test between the observationally derived WDLF and the WDLFs
  computed under different anomalous energy losses.}
\label{Fig:WDLFs}
\end{figure}

\section{Discussion and conclusions}
As shown by \cite{2014A&A...562A.123M} different modern WDLFs do not agree
with each other within their own quoted error bars. Then, a conservative
estimation of the WDLF of the Galactic disk and its uncertainties can be done
by taking two completely independent derivations of the WDLF and estimating
the differences between both (see \citealt{2014A&A...562A.123M} for
details). In Fig. \ref{Fig:WDLFs} we compare the theoretical WDLFs with the
WDLF of the Galactic disk constructed in this way. It is worth noting that, in
agreement with \cite{2008ApJ...682L.109I}, the upper panel of
Fig. \ref{Fig:WDLFs} shows that in the case of axions the best fit is obtained
for DFSZ-axions masses of $\sim 5$meV. While this result is only marginally
significant from a statistical perspective (see inset) it is interesting in
the light of the forthcoming International Axion Observatory (IAXO) which will
be able to explore axion masses in the range $m_a\cos^2\beta\gtrsim 3$ meV. As
shown in the insets of Fig. \ref{Fig:WDLFs}, a $\chi^2$-test indicates that
WDLFs constructed with values of $g_{ae}\gtrsim 2.3\times 10^{-13}$ and
$\mu_\nu \gtrsim 5 \times 10^{−12}e\hbar/(2mec)$ are at variance with the
WDLF of the Galactic disk at the $2\sigma$-level. These values are close to
the best available constraints coming from the study of globular clusters
\citep{2013PhRvL.111w1301V}. While this shows that modern WDLFs are an
excellent tool for constraining the properties of axions and neutrinos these
constraints should not be taken at face value. First, theoretical
uncertainties in the theoretical cooling times might be as high as $\sim 10$\%
in some regimes \citep{2010ApJ...716.1241S}. Second, departures form the
constant stellar formation rate, as those inferred by
\cite{2013MNRAS.434.1549R}, might also introduce uncertainties of the order of
$\sim 10$\% in the computed values. When taken into account, these two effects
should slightly weaken the previous constraints. Even more, discrepancies
between different WDLFs suggest there might be some relevant unaccounted
systematic errors. A larger set of completely independent WDLFs, as well as
more detailed studies of the theoretical WDLFs, and their own uncertainties,
is desirable to explore the systematic uncertainties behind these constraints.



\begin{thebibliography}{}
\expandafter\ifx\csname natexlab\endcsname\relax\def\natexlab#1{#1}\fi
\expandafter\ifx\csname url\endcsname\relax
  \def\url#1{\texttt{#1}}\fi
\expandafter\ifx\csname urlprefix\endcsname\relax\def\urlprefix{URL }\fi
\providecommand{\eprint}[2][]{\url{#2}}

\bibitem[{{Blinnikov} \& {Dunina-Barkovskaya}(1994)}]{1994MNRAS.266..289B}
{Blinnikov}, S.~I., \& {Dunina-Barkovskaya}, N.~V. 1994, \mnras, 266, 289

\bibitem[{{Dine} et~al.(1981){Dine}, {Fischler}, \&
  {Srednicki}}]{1981PhLB..104..199D}
{Dine}, M., {Fischler}, W., \& {Srednicki}, M. 1981, Physics Letters B, 104,
  199

\bibitem[{{Harris} et~al.(2006){Harris}, {Munn}, {Kilic}, {Liebert},
  {Williams}, {von Hippel}, {Levine}, {Monet}, {Eisenstein}, {Kleinman},
  {Metcalfe}, {Nitta}, {Winget}, {Brinkmann}, {Fukugita}, {Knapp}, {Lupton},
  {Smith}, \& {Schneider}}]{2006AJ....131..571H}
{Harris}, H.~C., {Munn}, J.~A., {Kilic}, M., et al. 2006, \aj, 131, 571.
  

\bibitem[{{Iben} \& {Laughlin}(1989)}]{1989ApJ...341..312I}
{Iben}, I., Jr., \& {Laughlin}, G. 1989, \apj, 341, 312

\bibitem[{{Isern} et~al.(2008){Isern}, {Garc{\'{\i}}a-Berro}, {Torres}, \&
  {Catal{\'a}n}}]{2008ApJ...682L.109I}
{Isern}, J., {Garc{\'{\i}}a-Berro}, E., {Torres}, S., \& {Catal{\'a}n}, S.
  2008, \apjl, 682, L109. 

\bibitem[{{Jaeckel} \& {Ringwald}(2010)}]{2010ARNPS..60..405J}
{Jaeckel}, J., \& {Ringwald}, A. 2010, Annual Review of Nuclear and Particle
  Science, 60, 405. 

\bibitem[{{Krzesinski} et~al.(2009){Krzesinski}, {Kleinman}, {Nitta},
  {H{\"u}gelmeyer}, {Dreizler}, {Liebert}, \& {Harris}}]{2009A&A...508..339K}
{Krzesinski}, J., {Kleinman}, S.~J., {Nitta}, A., {H{\"u}gelmeyer}, S.,
  {Dreizler}, S., {Liebert}, J., \& {Harris}, H. 2009, \aap, 508, 339

\bibitem[{{Miller Bertolami}(2014)}]{2014A&A...562A.123M}
{Miller Bertolami}, M.~M. 2014, \aap, 562, A123

\bibitem[{{Miller Bertolami} et~al.(2014){Miller Bertolami}, {Melendez},
  {Althaus}, \& {Isern}}]{2014arXiv1406.7712M}
{Miller Bertolami}, M.~M., {Melendez}, B.~E., {Althaus}, L.~G., \& {Isern}, J.
  2014, ArXiv e-prints. \eprint{1406.7712}

\bibitem[{{Raffelt}(1996)}]{1996slfp.book.....R}
{Raffelt}, G.~G. 1996, {Stars as laboratories for fundamental physics}
  (University of Chicago Press)

\bibitem[{{Renedo} et~al.(2010){Renedo}, {Althaus}, {Miller Bertolami},
  {Romero}, {C{\'o}rsico}, {Rohrmann}, \&
  {Garc{\'{\i}}a-Berro}}]{2010ApJ...717..183R}
{Renedo}, I., {Althaus}, L.~G., {Miller Bertolami}, M.~M., {Romero}, A.~D.,
  {C{\'o}rsico}, A.~H., {Rohrmann}, R.~D., \& {Garc{\'{\i}}a-Berro}, E. 2010,
  \apj, 717, 183. 

\bibitem[{{Rowell}(2013)}]{2013MNRAS.434.1549R}
{Rowell}, N. 2013, \mnras, 434, 1549. 

\bibitem[{{Rowell} \& {Hambly}(2011)}]{2011MNRAS.417...93R}
{Rowell}, N., \& {Hambly}, N.~C. 2011, \mnras, 417, 93

\bibitem[{{Salaris} et~al.(2010){Salaris}, {Cassisi}, {Pietrinferni},
  {Kowalski}, \& {Isern}}]{2010ApJ...716.1241S}
{Salaris}, M., {Cassisi}, S., {Pietrinferni}, A., {Kowalski}, P.~M., \&
  {Isern}, J. 2010, \apj, 716, 1241.

\bibitem[{{Salaris} et~al.(2009){Salaris}, {Serenelli}, {Weiss}, \& {Miller
  Bertolami}}]{2009ApJ...692.1013S}
{Salaris}, M., {Serenelli}, A., {Weiss}, A., \& {Miller Bertolami}, M. 2009,
  \apj, 692, 1013. \eprint{0807.3567}

\bibitem[{{Viaux} et~al.(2013){Viaux}, {Catelan}, {Stetson}, {Raffelt},
  {Redondo}, {Valcarce}, \& {Weiss}}]{2013PhRvL.111w1301V}
{Viaux}, N., {Catelan}, M., {Stetson}, P.~B., {Raffelt}, G.~G., {Redondo}, J.,
  {Valcarce}, A.~A.~R., \& {Weiss}, A. 2013, Physical Review Letters, 111,
  231301. 

\bibitem[{{Zhitnitsky}(1980)}]{Zhitnitsky}
{Zhitnitsky}, A.~R. 1980, Sov. J. Nucl. Phys., 31, 260

\end{thebibliography}


\end{document}